\documentstyle[aps,manuscript]{revtex}  
\begin{document}
\draft
\title { Metrical Separability of the Spin Systems
 Energetic Surfaces and Quantum Jumps Hypothesis }

\author{V.A.Skrebnev and M.N.Ovchinnikov}

\address{Faculty of Physics, University of Kazan,
Kremlevskaya 18, 420008, Kazan, Russia}

\date{\today}

\maketitle
\begin{abstract}

Metrical separability of the spin systems energetic
surfaces is shown. The irreversibility of the spin
system evolution  is considered as the consequence of
random quantum jumps on energy surfaces.
\end{abstract}


\narrowtext

It is known that infinitely great time is necessary
for the macrosystem to search through all their
possible mechanical states. It means that during
the measurement time the macrosystem can cover
only insignificant small part of these states.
In this connection it seems to be highly strange,
that the commonly accepted procedure of averaging
over all states leads to the agreement of a calculated
result with an experimental data.

The averaging over the phase space is possible
only in the case, when the system can reach all its states. Thus, the
demand of the phase space non-separability lies in a foundation of the
statistical physics, though the correctness of this demand is not proved
in the framework of mechanics. Usually the authors of wellknown books on
statistical mechanics refer to intuition at the discussion of this
question.

We will show  on an extremely simple example, that
at least for spin systems the demand of phase space non-separability does
not fulfill. 
   
As it will be disclosed below  the averaging over the phase space becomes
feasible if we add spontaneous transitions or quantum jumps on the
isoenergetic surfaces to the usual quantum-mechanical
description of the system evolution and will consider the system
state as a superposition of the states arrived as the result of such
jumps.

Consider the system of three interacting spins $I=1/2$ with Hamiltonian
\begin{eqnarray}
  H=H_z+H_d+P
\label{f1}
\end{eqnarray}

 where

 \begin{equation}
    H_z=\omega I_z=\omega \sum_{i} I_z^i
\end{equation}
\begin{equation}
H_d=\sum_{i<j} a [I_z^iI_z^j-\frac{1}{4}(I_+^i I_-^j+I_-^i I_+^j)]
\label{f3}
\end{equation}
\begin{equation}
   P= \sum_{i<j} a [I_+^i I_+^j+I_-^i I_-^j]
     \label{f4}
\end{equation}

Let's choose functions as basis functions the wave functions

\begin{equation}
  \Phi_k= \prod_i \varphi^i_l
\label{f5}
\end{equation}

where $  \varphi^i_l  $  are the eigenfunctions of operator
$I_z^i$  .

For the system of three spins
these functions are:
\begin{eqnarray}
 \Phi_1=\varphi^{(1)}_{\frac{1}{2}}    \varphi^{(2)}_{\frac{1}{2}}
\varphi^{(3)}_{\frac{1}{2}};
 \Phi_2=\varphi^{(1)}_{-\frac{1}{2}}    \varphi^{(2)}_{\frac{1}{2}}
\varphi^{(3)}_{\frac{1}{2}};
\nonumber\\
\Phi_3=\varphi^{(1)}_{\frac{1}{2}}    \varphi^{(2)}_{-\frac{1}{2}}
\varphi^{(3)}_{\frac{1}{2}};
 \Phi_4=\varphi^{(1)}_{-\frac{1}{2}}    \varphi^{(2)}_{-\frac{1}{2}}
\varphi^{(3)}_{\frac{1}{2}};
\nonumber\\
\Phi_5=\varphi^{(1)}_{\frac{1}{2}}    \varphi^{(2)}_{\frac{1}{2}}
\varphi^{(3)}_{-\frac{1}{2}};
 \Phi_6=\varphi^{(1)}_{-\frac{1}{2}}    \varphi^{(2)}_{\frac{1}{2}}
\varphi^{(3)}_{-\frac{1}{2}};
\nonumber\\
\Phi_7=\varphi^{(1)}_{\frac{1}{2}}    \varphi^{(2)}_{-\frac{1}{2}}
\varphi^{(3)}_{-\frac{1}{2}};
 \Phi_8=\varphi^{(1)}_{-\frac{1}{2}}    \varphi^{(2)}_{-\frac{1}{2}}
\varphi^{(3)}_{-\frac{1}{2}}.
\label{f6}
\end{eqnarray}

The arbitrary function of the system state can be expanded over
the functions of the basis (\ref{f5}) as

\begin{equation}
   \Psi=\sum_n C_{n}(t) \Phi_n
     \label{f7}
\end{equation}

Being substituted into  Schr$\ddot o$dinger equation it gives the
equation for coefficients $ C_n(t) $
\begin{equation}
  i \frac{dC_{m}(t)}{dt}=\sum_n C_{n}(t) H_{nm}
\label{f8}
\end{equation}

where $H_{mn}$ is the  matrix  of  Hamiltonian  (\ref{f1})  on  functions
(\ref{f5}).  The space of coefficients $ C_n$ ($C$-space) is analogous
to the phase space in classical mechanics.

         Computers make it easy to follow the evolution of the system
under consideration at various initial conditions. The system evolution
was examined at the following initial values of coefficients $C_n$ :
$C_2(0)=1$, others $ C_i(0)=0.$ The numerical
integration was made by using a fourth-order Runge-Kutta algorithm.
In experiments the stability of equation (\ref{f8}) was manifested, the
total energy was conserved and the normalization requirement was
fulfilled.  FIG.1 demonstrate the evolution of  value
 of $I_z$   operator  for  every  spin \\ $ <I_z^i>=  <\Psi^ *(t)\vert
I_z^i(t)\vert \Psi (t)>$ \ \ at
$\omega=10, a=1$.  It could seem at first that the mutual re-orientation
of the spins,  which  takes  place  because  of  the  interactions
(\ref{f3})
would result in transitions of the system into the state
$\Phi_3$ and $\Phi_5$. The consequence of such transitions would be identical
behaviour of $<I_z^i> $ on large enough observation times. It is seen
however that the values $<I_z^i>$ are changing periodically and remain
essentially different. The mean values$ <I_z^i>_{av}$ on a
large observation time prove to be following

$$
<I_z^1>_{av}=-0.06,  <I_z^2>_{av}=0.27, <I_z^3>_{av}=0.27
$$

The surface $ E_c$ in the space $C_n$  corresponds to a certain
value of system energy. The calculations have demonstrated
that the system cannot move from the     initial state
$\Phi_2$ to state $\Phi_3$ or $\Phi_5$, though they are equivalent from
the energetic point of
view.  It means that the states $\Phi_2, \Phi_3$ and $\Phi_5$ belong to
different $E_c$-surfaces. We also investigated the behaviour of
the spin systems, consisting of 4, 6 and 8 interacting particles.
The results turned out to be analogous to those for the system of
three spins. Thus in the $C$-space there exist some non-intersecting
surfaces $E_c$ with the same energy value. This fact leads to the
essential, in our opinion, conclusion about metrical separability
of the $ E_c$-surface.

Let's assume  now  the  existence  of  random quantum jumps on the
surface $E_c$
and between different $E_c$-surfaces with the same    energy value.
In consequence of these jumps  the states, which are equal from the
energy point of view turn out also to be equally probable.
Thus such random jumps make   it possible to eliminate
the thermodynamically inconvenient consequences of the $ C$-space metric
separability.

  We would like to note    that the influence of the quantum jumps as
the natural process in nature upon the evolution of physical systems
is considered  in  a  number   of   works   (see,   for   example,
\cite{l1,l2}).
In \cite{l1}
the geometrical and temporal conditions of the possible wave
functions collapse of the microscopical
quantum system are considered. The frequency of the quantum jumps
has no considerable meaning for our problem. It is however evident
that this frequency should be low enough. In \cite{l1,l2} the
 probability of the quantum jump for one particle is assumed as
$10^{-16}$. In other words, the collaps occurs in $10^8$ years. The low
probability of the quantum jumps determines the high  precision
of the Schr$\ddot o$dinger equation for the systems with the
small number of particles. At the same time, it is evident that
quantum jumps can turn out to be very essential for the system
with number of particles $10^{23}$ .

The random quantum jumps on and between $E_c$-surfaces containing
the states $\Phi_2 , \Phi_3$ and $\Phi_5$  bring to identical mean values of
spins Zeeman energy $<I_z^i>_{av}=0.16$.

         The equality of the mean values of
$<I_z^i>_{av}$  allows to present them  as:

\begin{equation}
    < I_z^i>_{av}=Tr(\hat \rho_i \hat I_z^i)
\label{f9}
\end{equation}

where
\begin{equation}
     \hat \rho_i=A \exp (-\beta \omega \hat I_z^i)
\label{f10}
\end{equation}

The value of $\beta$ is determined by initial conditions and
is equal for all spins.

It follows from (\ref{f9}) that:

\begin{equation}
     <I_z^i>_{av}=-\frac{1}{2} \tanh \frac{\beta \omega}{2}
 \label{f11}
\end{equation}

Thus, spontaneous transitions on the energy surfaces considered
as an addition to the quantum mechanical evolution provide the
density matrix factorization and transfer to statistical
description of the system.

It is difficult to imagine the experiments which would
allow to detect the manifestation of random quantum jumps
in the systems with small number of particles. The case
is essentially different for spin macrosystems. It has
been demonstrated in the works \cite{l3,l4} that the results
of the experiments on
observing the establishment of equilibrium in the  spin
systems cannot be described by the reversible equations
of quantum mechanics.

The equations like (\ref{f8}) do not allow  the exponential
divergence of solutions. As a result, the attempts to
use the dynamic chaos ideology to explain the real
irreversibility in the quantum system evolution cannot
be fruitful. It should be noted that
in the classical case also the instability of reversible
mechanics equations does not bring the system in the state
with maximum entropy. It seems naive to explain the  
existence of real chaos in system as a result of influence
of environment,  where this real chaos arises by mysterious way.

A natural explanation for the actual irreversibility
discovered in the works \cite{l3,l4,l5} can be provided by the existence
of the  random quantum jumps between the  states with
equal energy.  The consequence of quantum jumps is the possibility
to consider the macroscopical state of the system as a superposition
of the equally probably states with equal energy. Averaging on
these states is equivalent to ones on statistical ensemble and
the correctness of such averaging did not demand infinite
large observation time.

\acknowledgements
The authors are grateful to prof. L.Tagirov for the fruitful discussions
and help with the work.

\newpage
\begin{figure}
\caption{ The    time    dependence    of    $<I_z^1>$ ( continuous line, curve
1),  $<I_z^2>$ ( dotted line, curve 2),  $<I_z^3>$ ( dotted line, curve 3)}
\end{figure}

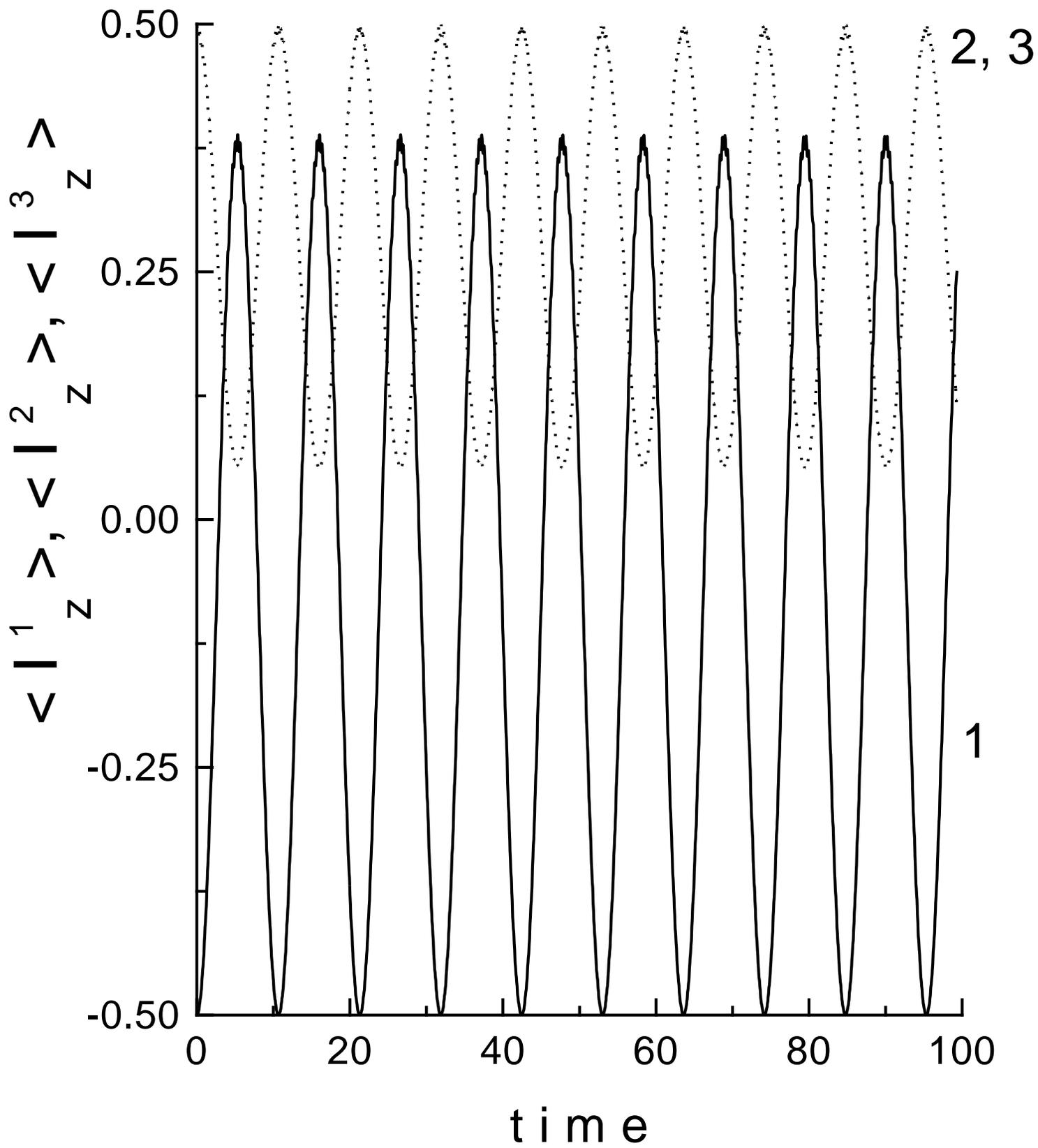
\end{document}